\documentclass[12pt]{article}

\usepackage{times}

\usepackage{setspace}

\usepackage{graphicx}
\usepackage{amsmath}
\usepackage{amsfonts}
\usepackage{amssymb}

\usepackage{xcolor}
\usepackage[normalem]{ulem}

\newcommand{\Alfven}{Alfv\'{e}n}
\newcommand{\sciexp}[2]{{#1}\ensuremath{\,\times\,10^{#2}}}

\newcommand{\eff}{{e\!f\!\!f}}

\newcommand{\um}{$\mu$m}

\newcommand{\PU}{Department of Astrophysical Sciences, Princeton University, Princeton, NJ 08544, USA}
\newcommand{\PPPL}{Princeton Plasma Physics Laboratory, Princeton, NJ 08543, USA}
\newcommand{\UMich}{Center for Ultrafast Optical Science, University of Michigan, Ann Arbor, MI 48109, USA}

\newcommand{\LLE}{Laboratory for Laser Energetics, University of Rochester, Rochester, New York 14623, USA}
\newcommand{\UNH}{Space Science Center, University of New Hampshire, Durham, New Hampshire 03824, USA}
\newcommand{\CU}{University of Colorado, Boulder, CO 80309, USA}

\title{A novel kinetic mechanism for the onset of fast magnetic reconnection and plasmoid instabilities in collisionless plasmas}

\author   {W.~Fox,$^{1,2,\ast}$, 
G.~Fiksel,$^{3}$
D.~B.~Schaeffer,$^{2}$
D.~Haberberger,$^{4}$
J.~Matteucci,$^{2}$ \\ 
K.~Lezhnin,$^{2}$ 
A.~Bhattacharjee,$^{1,2}$
M.J.~Rosenberg,$^{4}$ 
S.X.~Hu,$^{4}$
A.~Howard,$^{4}$ \\
D.~Uzdensky,$^{5}$
K.~Germaschewski$^{6}$ \\
\normalsize{$^{1}$ \PPPL} \\
\normalsize{$^{2}$ \PU}  \\
\normalsize{$^{2}$ \UMich}  \\
\normalsize{$^{4}$ \LLE} \\
\normalsize{$^{5}$ \CU} \\
\normalsize{$^{6}$ \UNH} \\
\\
\normalsize{$^{\ast}$To whom correspondence should be addressed; E-mail:  wfox@pppl.gov }
}

\begin{document}

\maketitle


\begin{abstract} 

Magnetic reconnection can explosively release magnetic energy when opposing magnetic 
fields merge and annihilate through a current sheet, driving plasma jets and 
accelerating non-thermal particle populations to high energy,
in plasmas ranging from space and astrophysical to laboratory scales 
\cite{ MasudaNature1994,YamadaRMP2010,MatsumotoScience2015,BurchScience2016}.
Through laboratory experiments and spacecraft observations,
significant experimental progress has been made in
demonstrating how fast dissipation and reconnection occurs in narrow, kinetic-scale 
current sheets \cite{BurchScience2016,RenPRL2005,FoxPRL2017}.
However, a challenge has been to demonstrate what triggers reconnection and
how it proceeds rapidly and efficiently as part of a global system much larger
than these kinetic scales. 
Here we show experimentally
the full development of a process where the current sheet  forms and then breaks up into multiple current-carrying structures
at the ion kinetic scale.
The results are consistent with tearing of the current
sheet, however modified by collisionless kinetic ion effects,
which leads to a larger growth rate and number of plasmoids than observed in previous experiments \cite{HarePRL2017} 
or compared to predictions from standard tearing instability theory \cite{BhattacharjeePoP2009} 
and previous non-linear kinetic reconnection simulations \cite{DaughtonNature2011, DaughtonPRL2009}.
This effect will increase the role of plasmoid instabilities in many natural reconnection
 systems and should be considered in triggering rapid reconnection 
in a broad range of natural plasmas with collisionless, compressible 
flows, including at the Earth's magnetosheath \cite{RetinoNatPhys2007, PhanNature2018} and
magnetotail \cite{ChenNature2008} and at the heliopause \cite{OpherApJ2011}, in accretion disks \cite{HoshinoPRL2015}, 
and in turbulent high-Mach-number collisionless shocks \cite{MatsumotoScience2015}.

\end{abstract}

\maketitle

\doublespacing

In weakly collisional plasmas, current sheets typically thin to kinetic scales comparable
to the ion skin depth $d_i = (m_i / \mu_0 n_e Z_{\eff} e^2)^{1/2}$
\cite{YamadaRMP2010, BurchScience2016, RenPRL2005, FoxPRL2017}.
In most astrophysical 
and space plasmas of interest, however, there is a very large separation of scales
between global scales driving reconnection and the kinetic scales where dissipation occurs,
which presents a challenge for understanding efficient conversion of magnetic field energy.
The large scale separation from the global scale to kinetic and resistive scales 
is indicated by large values of the dimensionless system size $L / d_i $ and the
Lundquist number $S = \mu_0 L V_A / \eta$, where $L$ is the current sheet
length, $V_A = B / \sqrt{\mu_0 n_i m_i}$ is the \Alfven{} speed, and $\eta$ the
plasma resistivity \cite{JiPoP2011}.
Theory and recent simulation have proposed how plasmoid instabilities 
\cite{BhattacharjeePoP2009, DaughtonNature2011, JiPoP2011, ShibataEPS2001, LoureiroPoP2007, UzdenskyPRL2010}
grow and interact to generate turbulent flows near reconnection
layers.  The resulting plasmoid turbulence simultaneously forces a large number 
of narrow current sheets for fast dissipation while maintaining
 broad outflow layers for efficient global energy conversion.
This scenario remains to be demonstrated in experiments, though the initial steps of this process are indicated
by laboratory observations of individual plasmoids 
\cite{HarePRL2017, GekelmanJGR1991,FikselPRL2014,OlsonPRL2016,JaraAlmontePRL2016},
or in spacecraft data which appears to cross such structures \cite{ChenNature2008}.
Experiments with high-energy laser-produced plasmas allow access
to reconnection in such a large-system size regime, with current sheets
much longer than both ion and electron kinetic scales, $L/d_i \sim 50$,
and at large Lundquist number ($S \sim 1000$) \cite{RosenbergPRL2015}
which can therefore address these questions \cite{FoxPRL2011}.
We note this normalized system size approaches that for the Earth's
magnetotail (where $L \sim 10 R_E \sim 100 d_i$, where $R_E$ is the Earth's radius).
Using this technique, in this Letter we present laboratory experiments 
with collisionless ions in large-system size regime showing 
the formation and subsequent breakup of a reconnection current sheet.
The results are consistent with tearing instabilities, though we observe larger growth rates,
and a larger number of structures than were 
produced in previous experiments \cite{HarePRL2017}, or predicted by classical tearing instability theory 
\cite{BhattacharjeePoP2009, CoppiPRL1966, BaalrudPoP2011} 
and previous non-linear kinetic reconnection simulations \cite{DaughtonPRL2009}.

\begin{figure*}

\includegraphics[width=6in]{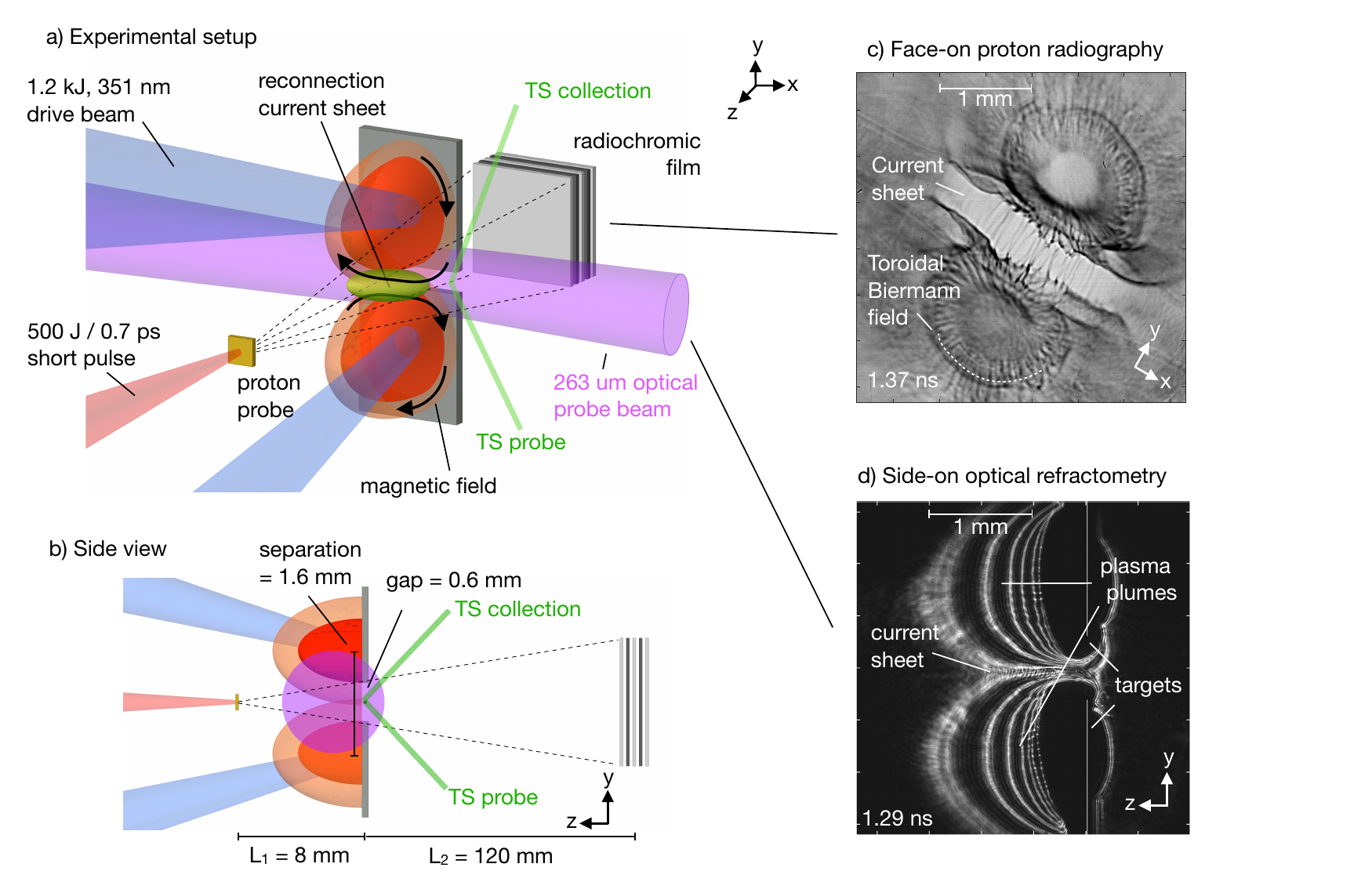}
\caption{\label{Fig_Overview} \textbf{Experimental setup for observations of
magnetic reconnection by proton radiography and optical refractometry.}  (a-b) The experiment consists of
two parallel plastic (CH) foils, each driven by a 1.25~kJ, 1~ns laser at 351~nm to produce
two expanding plasma plumes.  
We define a coordinate system where $x$ is in the plane of the foil and
corresponds to the reconnection outflow direction, $y$ is in the plane of the foil corresponding
to the reconnection inflow direction, and $z$ is perpendicular to the foil corresponding to the
direction of plasma current flow in the current sheet.
The two foils were separated by a gap of
600~\um.  Thomson scattering is used to measure plasma density and separate electron and ion temperatures.  The
Thomson scattering volume was in the plane of the foils at a position $y=0.8$~mm
``downstream'' of the collision midpoint between the two plumes.
(c) Proton radiography was obtained using a beam of fast protons 
produced by irradiating a Cu-foil with a short-pulse laser
(480 J in 0.7~ps), which produced a high-time- ($\sim$~ps)
and space-resolution ($<20~\mu$m) point source of 
protons with a broad energy distribution
out to 10's of MeV.  The protons stream through the interaction region and deposit in a radiochromic film (RCF)
stack, which provides energy resolution.  The protons probed the plasma along the $z$-axis, yielding
proton radiography images in the $x$-$y$ plane.
An example film image is shown corresponding to 30~MeV protons.
(d) Side-on optical probing was obtained with a 263~nm probe beam connected to an angular-filter
refractometry (AFR) system.  The beam propagated along the $x$-axis
yielding images in the $y-z$ plane corresponding to density line-integrated in $x$.  
The bands map contours of constant 
gradient in line-averaged density,
giving the global plume structure (Supplementary).
The initial locations of the target foils are indicated by the gray lines.
}
\end{figure*}

Experiments at the OMEGA EP Laser Facility were used to
collide and merge two magnetized plasma plumes
and drive reconnection (Fig.~\ref{Fig_Overview}).
Following the techniques of previous laser-driven reconnection
experiments \cite{NilsonPRL2006, RosenbergPRL2015, FikselJPP2021},
a pair of expanding, interacting plasma plumes
are generated with two laser pulses irradiating a pair of
thin CH foils. As the plumes expand, they each self-generate a strong toroidal magnetic field ($B =~$20-40~T)
by the Biermann battery effect, and, as the two plasmas
merge together, these fields interact and reconnect.
In these experiments we introduce a gap between two adjacent foils
to allow diagnostic 
access for high resolution proton radiography, optical refractometry, and Thomson scattering
of the reconnection layer  (Fig.~\ref{Fig_Overview}a,b).
The colliding ablated flows compress the anti-parallel
fields, leading to a  strongly-driven magnetic reconnection in a compressible regime relevant to 
high Mach number shocks and other compressible flows in heliophysics and astrophysics.
The plasma density and temperature is measured
by Thomson scattering and ranges from $n_e =$~1--3~$\times~10^{25}$~m$^{-3}$, with $T_e =$~400--600~eV and $T_i =$~1600--2400~eV.
The corresponding plasma $\beta_e = 2 \mu_0 n_e T_e / B^2 =$~2--8 and dynamic $\beta_{\textrm{dyn}} = \mu_0 n_i m_i V_{in}^2 / B^2=~$~2.5--10,
assuming inflows $V_{in} \approx$~250~km/s
at the time of reconnection.
The ion skin depth $d_i \approx$~60--100~$\mu$m demonstrates the large separation
of scales between kinetic scales and the global length of the current sheet, $L_{CS} \sim 4$~mm~= ~40--80~$d_i$.
The high plasma temperature implies a long ion mean-free-path 
($\lambda_{ii} > $~0.5~mm,
between the counterflowing plasmas) and a high Lundquist number $S \sim$~2500--4500.  (Methods of
establishing plasma parameters are discussed
in the Supplementary and presented in Table~\ref{Tab_Param}).

Imaging data (Fig.~\ref{Fig_Overview}c and d) shows the
evolution of the magnetic fields and plasma density in the colliding plasmas.
The proton radiography data (Fig.~\ref{Fig_Overview}c) is obtained using a beam of high-energy protons,
which streamed through the merging plasmas,
and were recorded  in a stack of radiochromic film.
The protons
experience small deflections from the transverse electric and magnetic fields
along their trajectories, so that the fluence variations
are related to the line-integrated electromagnetic fields \cite{BottJPP2017}.
For this geometry, protons respond
to line-integrated magnetic field components
($\int B_x dz$ and $\int B_y dz$), 
which correspond to the reconnection 
inflow and outflow magnetic field
components.
Secondly, at these plasma parameters and
proton-probing energies, the protons
respond predominantly to magnetic rather than
electric fields \cite{PetrassoPRL2009}.
The radiographs are analyzed quantitatively below, but it is useful to first point
out the qualitative features in the data.
First, the thin dark circular feature, marked ``Toroidal Biermann field'', is due to the 
proton focusing by the global toroidal field by the global Biermann battery effect
in each plume.
Where the plumes begin to merge and interact, a current
sheet is formed, observed as a broad ($\sim 450$~$\mu$m)
fluence depletion (light, marked ``Current sheet''), bounded by two
narrow fluence enhancements (dark), similar to observations in Ref.~\cite{RosenbergPRL2015}.
The fluence depletion is 
consistent with the defocusing of protons by the 
reversing magnetic field across the current sheet.
Transverse fluence variations are also observed along the sheet, as well
as around the plumes, related to small scale magnetic fluctuations.
Side-on optical refractometry is simultaneously obtained
using a 263~nm optical probe beam (Fig.~\ref{Fig_Overview}d) to observe
 the plasma density.  
Significant plasma interaction is observed
where the two plasmas meet and form the current sheet.

\begin{figure*}
\centering
\includegraphics[width=6.5in]{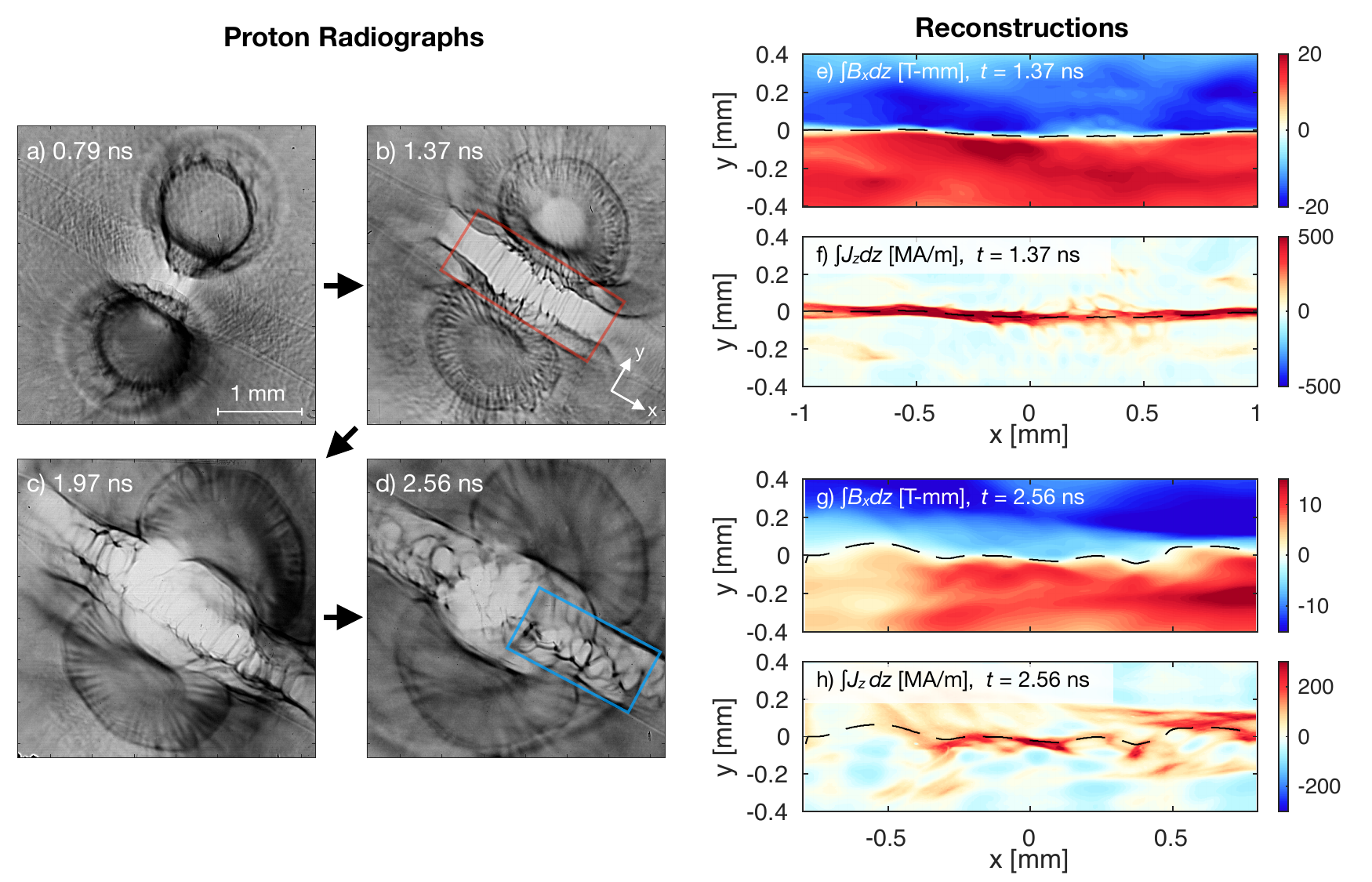}
\caption{\label{Fig_Prad}
\textbf{Proton radiography observations of current sheet formation and breakup.}
(a-d) Raw proton radiography data images at four times show the collision
of two plumes.
The images were obtained on separate shots, and 
therefore individual features cannot be tracked between frames, however
the global features including the 
formation of an extended current sheet, 
and subsequent current-sheet breakup is apparent.
The data correspond to film in the RCF stack corresponding to 30~MeV protons.  
(e,f) Reconstructions of the reconnecting magnetic field $(\int B_x dz$, panel e) and plasma current
density ($\int J_z dz$, panel f) corresponding to the magenta box 
in the early-time (data 1.37~ns, panel b).
(g,h) Same, corresponding to the blue box in the late time data (2.56 ns, panel d).
The dashed line indicates the magnetic reversal surface, where $\int B_x dz = 0$.
The images (a-d) have been sharpened and contrast-adjusted, while data in 
(e-h) rely on processed raw data.  The 
processing and reconstruction methodology is described in detail in the Supplementary.}
\end{figure*}

A sequence of proton radiographs (Fig.~\ref{Fig_Prad})  shows the full evolution
of the interaction of the plumes, and formation and breakup of the current sheet. 
We present both the raw proton radiograph data (a-d), which contains valuable
qualitative information, as well as quantitative reconstructions of the magnetic field
$(\int B_x dz)$ and current ($\int J_z dz$) (e-h).
Soon after the two plasmas meet (Fig.~\ref{Fig_Prad}a, 0.79~ns), the 
current sheet region of reversing magnetic field 
forms, which quickly extends to nearly the full 3.5~mm field-of-view (by Fig.~\ref{Fig_Prad}c, 1.97 ns).
At early time, as the current sheet is forming, it is relatively uniform along its length (Fig.~\ref{Fig_Prad}b, 1.37 ns)
with relatively small current fluctuations evident in fine transverse proton
modulations.
However, by later time (Fig.~\ref{Fig_Prad}c and d, 1.97 and 2.56 ns),
the transverse modulations transition to a much higher-contrast 
with a cellular morphology.  The amplitude of these structures is now very high,
corresponding to $\sim$100\% variations in the proton fluence.

A reconstruction of the plasma current sheet quantitatively demonstrates that these
high-amplitude cellular structures correspond to a breakup of the current sheet.
The reconstruction procedure \cite{BottJPP2017} inverts the proton fluence maps to obtain the line-integrated magnetic fields and associated 
current density, where the line-integration is along
the trajectory (primarily $\hat{z}$) of the probing protons.
Figure~\ref{Fig_Prad}e and f
show the line-integrated magnetic field
and plasma current density at the 
early time $t=1.37$~ns as the current sheet is forming.
At this time, the current sheet is highly-extended and quasi-laminar, with
 half-width $a \equiv B_{up} / (dB/dy) = $30--40~$\mu$m.  The
upstream field $|\int B_x dz|$ is approximately 20~T-mm, and the
peak line-integrated current density ($\int J_z dz$) is near 400 MA/m.  Outside the current
sheet, the current density is slightly negative, reflecting the closure of the 
current back through the plasma plumes.

The reconstruction at later time (2.56~ns, Fig.~\ref{Fig_Prad}g and h) shows 
an evolution to a current sheet full of cellular structures.
We focus reconstruction
on a downstream region, centered approximately 0.8~mm from the midpoint and
indicated by the blue box of Fig.~\ref{Fig_Prad}d,
as this is the region which shows significant current sheet dynamics. 
The $B_x = 0$ surface, marked with a dashed line, has 
become highly kinked and folded.  The upstream fields have also 
weakened somewhat to $|\int B_{up}dz| \sim$~15~T-mm).
Most importantly, the 
current density has become highly modulated, with near 100\% modulations
in the current density.  
The typical scale of the current modulation is  200~$\mu$m.

\begin{figure}
\centering
\includegraphics{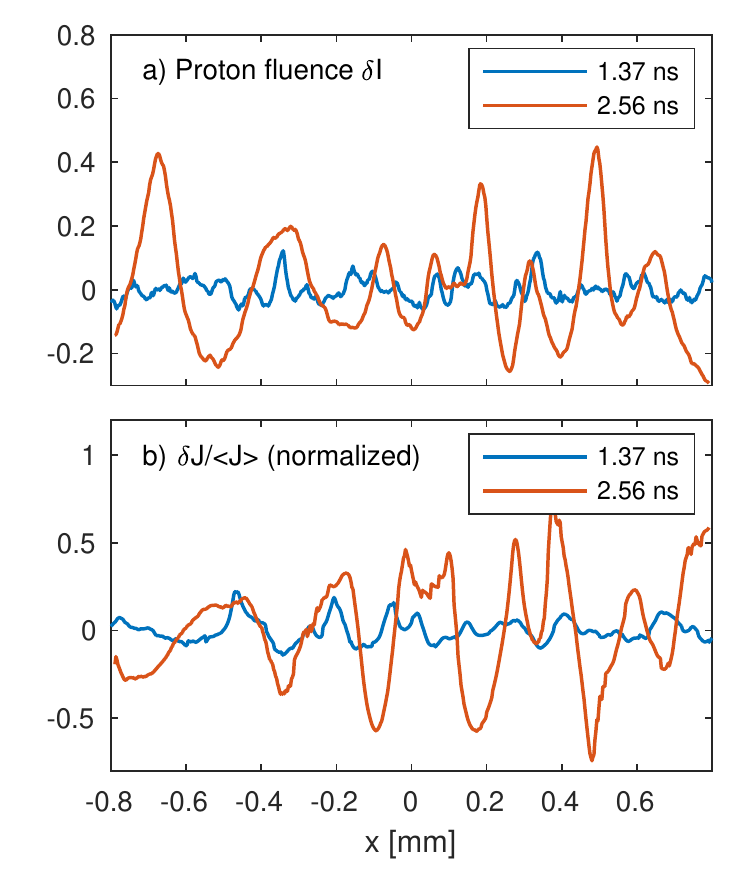}
\caption{\label{Fig_Prad_Zoom}
\textbf{Profiles of proton fluence and current density along the current sheet.}
a) Proton fluence cuts along the center of the current 
sheet ($y=0$) at early (1.37~ns) and late (2.56~ns).  
b) Corresponding reconstructed current density  ($\int J_z dz$) profiles
following the current sheet structure, normalized to the
average along the current sheet.
Both have been filtered to remove the mean and long-scale variations ($\lambda > 500$~um) to 
focus on the variations.}
\end{figure}

We examine profiles of the raw
proton and reconstructed current density fluctuations in Fig.~\ref{Fig_Prad_Zoom}, corresponding
to profiles taken along $B_x = 0$.
For each, we filter to isolate modulations with wavelength $< 500$~$\mu$m, and
for the current we normalize to the average current in the current sheet.
Both the protons fluctuation ($\delta I$) and current density fluctuations 
($\delta J/J$) increase
significantly, with the RMS increasing by a factor of $\sim$4 between the
two time points.
The data show a strong relation between 
raw proton fluence and plasma current. 
This is important as this confirms the relation
between high-contrast cellular proton fluence structures in the raw
data (Fig.~\ref{Fig_Prad}d) with the breakup of the current structures (Fig.~\ref{Fig_Prad}h).

Given that the proton measurement is line-integrating, we examined
whether the late-time data indicate that current sheet \textit{itself} has been folded and broken up, or whether
it can be explained by the superposition of a laminar current sheet with another independent, modulated magnetic field created by another process.
The evidence that the current sheet itself has been strongly folded and broken up is that 
the late-time data no longer shows any amount of the relatively clean early-time (1.37~ns) current sheet.
The key question is therefore to examine if current sheet instabilities can explain 
the breakup of the current sheet into multiple current carrying structures.
Taking the current modulation scale as 
indicative of the instability, and peaking at wavelengths $\lambda = 2\pi/k$ near 200~$\mu$m,
we evaluate $ka \approx 1$, where the current
sheet half-width $a = 30-40~\mu$m is obtained from the early-time magnetic profiles.
Based on the growth of magnetic fluctuations (Fig.~\ref{Fig_Prad_Zoom}, the instability growth rate $\gamma$ is in excess of $10^9$~s$^{-1}$, but
we note is likely fully into the non-linear regimes at late times
when the current sheet is nearly entirely broken up.
A natural mechanism to drive the breakup is the tearing instability.
However, the wavelength is smaller than expected for 
classical tearing, which is stabilized for $ka \gtrsim 1$.
We evaluated classical (resistive) \cite{FKRPoF1963, BhattacharjeePoP2009} 
and collisionless \cite{CoppiPRL1966} tearing for 
experimental parameters (Fig.~\ref{Fig_TearingGrowth}),
which both predict linear growth rates too slow to explain these
dynamics as well as fastest-growing  wavelengths longer than observed.
 For this reason, it is also perhaps not surprising that the
 typical wavelength 
 is also significantly shorter than observed
in recent nonlinear kinetic particle-in-cell simulations initiated from a Harris sheet \cite{DaughtonPRL2009} or under laser-plasma conditions \cite{LezhninPoP2018}.

We find the growth at shorter-wavelength may be explained by coupling of tearing to plasma 
temperature anisotropy, so-called anisotropic tearing, which was proposed previously as a mechanism to trigger
reconnection in the magnetotail \cite{ChenPoF1984, BurkhartPRL1989, QuestPoP2010}.
The anisotropy couples a mirror- (or Weibel-type) instability to the 
tearing dynamics and boosts growth rates and produces current-sheet-breakup at shorter wavelengths.
Phase-space anisotropy of the ions can develop from counterstreaming ions
or possibly compressive perpendicular heating resulting from the strong inflows,
parameterized by $\alpha = T_{\perp,\eff} / T_{||} - 1$, where
 $T_{\perp,eff} = M_i V_{in}^2 + T_\perp$,
and where the directions for $T_\perp$ and $T_{||}$ are with respect to the upstream field.

For the present experiments, using $t = 2$~ns, we estimate an anisotropy up to  $\alpha~\approx~$4
forms in C$^{6+}$ populations due to the strong plasma inflow, $V_{in}\sim$~250~km/s, compared to the
measured $T_{||}~\sim~$2~keV\@.
Figure~\ref{Fig_TearingGrowth} shows a calculation of growth rate 
versus wavelength for various
tearing models, including anisotropic tearing for several values of anisotropy.
Whereas classical (isotropic) tearing is stable at $\lambda = 200~\mu$m,
ion anisotropy boosts the growth rates and shifts the fastest-growing
wavenumbers to $ka =~$1--2 \cite{QuestPoP2010}, which can therefore explain the 
observations.
(While challenging to estimate due to their collisionality, anisotropy developing in the electron
populations can also strongly boost growth rates \cite{QuestPoP2010}.)
Interestingly, anisotropic tearing also
has a much larger growth rate than a pure Weibel instability (at $B=0$, calculated
in Fig.~\ref{Fig_TearingGrowth}, 
also pointed out by Ref.~\cite{QuestPoP2010}), owing to the presence of the
current sheet which magnetizes the electrons which allows the mode to grow more rapidly.
While pure ion-type Weibel instability was observed in previous
laser experiments \cite{FoxPRL2013, HuntingtonNatPhys2015}, the present 
observations are at finite magnetic field and the morphology
is apparently different, leading to closed cell magnetic structures rather than
extended filaments.

The observations motivate particle-in-cell plasma simulations 
to probe the interaction between these instabilities and 
reconnection.
We use the particle-in-cell simulation code PSC initialized with
plasma flow, density, and temperature profiles
obtained from a DRACO radiation-hydrodynamics
simulation, benchmarked against the observed density
evolution, and with magnetic fields consistent with the 
observed fields (Fig.~\ref{Fig_Prad}f, the setup and codes are discussed in the Supplementary).
Figure~\ref{Fig_Sim} shows a full 3-D visualization from the simulation of the 
structures formed during reconnection in the current sheet.
These large-scale simulations keep the full system size
in the inflow direction ($L / d_{i0} \approx 100$). 
In systems of this size, 
during the compression phase, current sheet is broken into a large number of plasmoids.
Synthetic proton radiography is obtained 
by ray-tracing protons from a point-source, through the simulation 
volume, and then projecting them ballistically to a detector plane,
where a fluence map is obtained.  The pattern obtained shows
the ``cellular'' structures which wrap around the O-points
formed by reconnection, showing how
 the change in magnetic topology and reconnection
 is manifested in the radiography.
 
\begin{figure}
\centering
\includegraphics[width=5in]{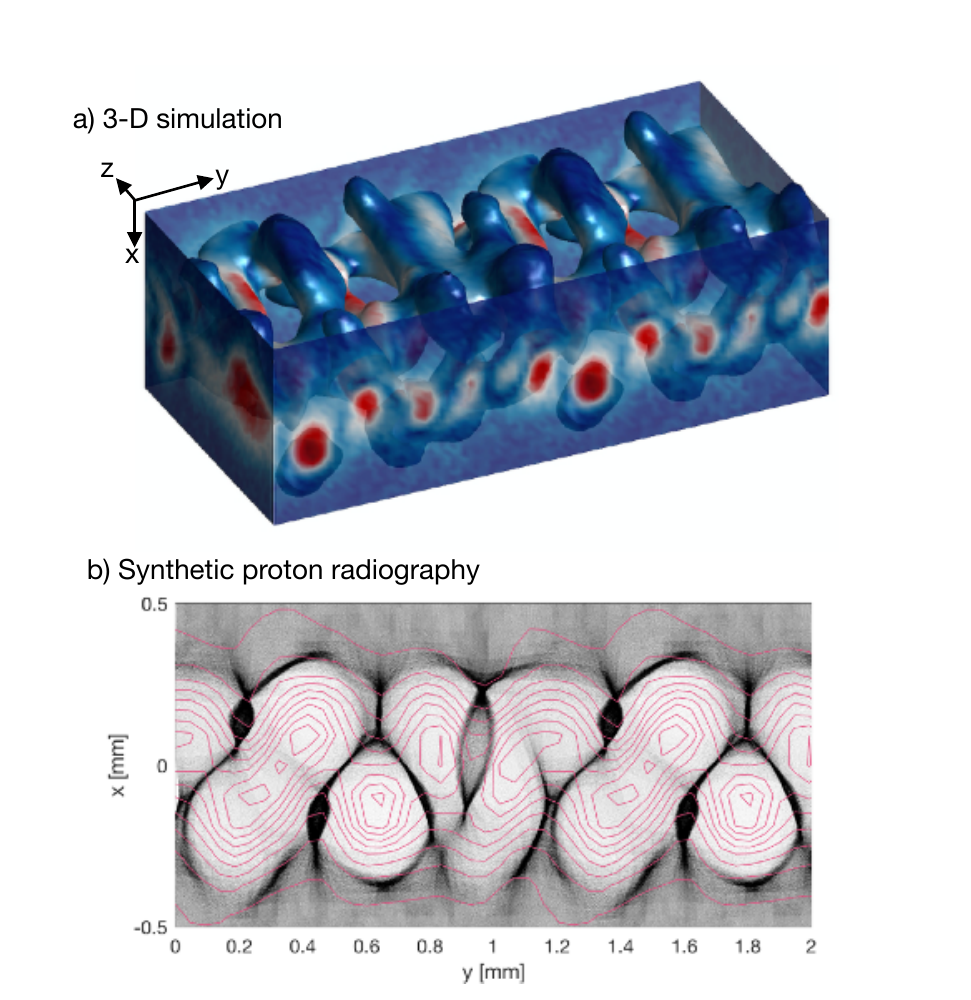}
\caption{\label{Fig_Sim} \textbf{Simulations of colliding
plasmas under experimental conditions.}
a) The 3-D visualization plasma current density (color) plotted
atop iso-contours of plasma density.  The face surfaces
display current density, showing the formation of
a large number of magnetic islands (high current density in red).
b) Synthetic proton radiography obtained from line-integrating
(along $z$) through the simulation fields, showing the development
of multiple cellular plasmoid structures from the reconnection of magnetic fields.
Following the conventions
of the experimental film, higher proton fluence is rendered as darker.
The magenta lines follow directly the simulation line-integrated magnetic fields $\int \mathbf{B}_{x,y} dz$,
and are included to show the connection between cellular structures in the 
proton radiography and change in the averaged magnetic topology.
Because of the 1-D nature of the plasma inflow initial condition used in the PIC simulation, a larger
plasma compression is observed than
in recent full 3-D simulations of similar systems, where
the extra dimensionality allows a transverse dilation of the plasma.
To account for this, we rescale the simulation length scales by matching the 
local ion-skin depth to the experiment. }
\end{figure}

These results show the full development
of the formation and breakup of a current sheet 
at a large-system size, high-Lundquist number and collisionless-ion regime.
The fast growth of instabilities at short wavelength ($ka \approx 1$) 
indicates the likely role of anisotropic tearing driven by collisionless ions.
These instabilities will boost reconnection rates and 
increase the number of plasmoids generated compared to pure tearing instability
theory,
which has implications for observational signatures and 
more broadly for enhancing reconnection in a number of
space plasma and astrophysical systems,
for example driving ``seas of plasmoids'' in the heliosheath \cite{OpherApJ2011, DrakeApJ2010}, and driving electron-scale turbulent reconnection
in the Earth magnetosphere \cite{PhanNature2018}.
These results demonstrate a rich physics connecting the
evolution of magnetic fields to the formation of 
non-thermal plasma distributions and to kinetic instabilities, 
important to energy conversion and transport processes 
in collisionless astrophysical and space plasmas,
which can now be studied in the laboratory, and may be observed in space
plasmas with rapid phase space measurements made
possible by the Magnetosphere Multiscale Mission \cite{BurchScience2016}.

\bibliography{refs}

\section{Acknowledgments}

The authors thank the OMEGA EP team for carrying out the experiments.
Experiment time was made possible by grant DE-NA0003613 provided by
the National Laser User Facility.
Simulations were conducted on the Titan supercomputer at the Oak Ridge Leadership Computing 
Facility at the Oak Ridge National Laboratory through the 
Innovative and Novel Computational Impact on Theory and Experiment (INCITE) program, which is
supported by the Office of Science of the DOE under Contract DE-AC05-00OR22725.
This research was also supported by the DOE under Contracts DE-SC0008655, DE-SC0008409, and DE-SC0016249,
and DE-AC02-09CH11466. 

\newpage

\appendix
\renewcommand{\thesection}{S}
\section{Supplementary Materials}

        \setcounter{table}{0}
        \renewcommand{\thetable}{S\arabic{table}}%
        \setcounter{figure}{0}
        \renewcommand{\thefigure}{S\arabic{figure}}%

\subsection{Detailed experimental setup}

Experiments were conducted at the OMEGA EP facility at the University of Rochester Laboratory for
Laser Energetics.
Two 6~mm $\times$ 4~mm plastic (CH) targets of thickness 25~$\mu$m were aligned in a common plane, separated
by a uniform gap of 600~$\mu$m (Fig.~\ref{Fig_Overview}a and b). 
The two main plasma plumes were each driven by a 1.25~kJ, 1~ns pulse 
of 351~nm ($3 \omega$) light, with foci separated across the gap by 1600~$\mu$m.  
Phase plates were used which produced two smoothed beam profiles with $1/e$~radius
of 340~um and
with a supergaussian intensity profile with index 8.  
The peak intensity was near \sciexp{3}{14}~W/cm$^2$.  
Over all shots analyzed from OMEGA EP in this study, the mean laser-energy-per plume was 1.25 kJ,
with the full range from 1.21 to 1.29~kJ, for a laser reproducibility of $\pm 3\%$.

A fast beam of probing protons was produced by a third, short-pulse IR ($1\omega$) 
laser, 480~J in 0.7~ps, focused to a spot size of $<25~\mu$m on a thin Cu foil.  
The thin Cu target was mounted 
in a backlighter assembly which is shielded from the expanding plasma with 
an additional thin Ta foil, all mounted in a 1~mm-diameter CH cylinder.
The short pulse produced a population of high energy probing protons with a broad distribution
of energies out to $>$~40~MeV by the target-normal sheath acceleration mechanism \cite{SnavelyPRL2000}.
Over several shots, the probing protons were variably timed with respect to the main drive plumes to 
radiograph different points of the plasma evolution and collision.

The probe protons pass through the plasmas of interest, and pick up deflections to their
trajectories by electric and magnetic field structures in the plasmas, after which
they propagate ballistically to a radiochromic film (RCF) stack.  The stack of 
film provides energy resolution due to the varying Bragg peak with proton energy.
For these experiments the proton source was $L_1$ = 8~mm from the foil plane, and detector
$L_2$~= 120~mm from the foil plane, for a geometrical magnification of $M=16$.
Distances in Figs.~\ref{Fig_Overview} and \ref{Fig_Prad} are given in
units in the plasma plane.  A full discussion of the 
processing and analysis of the proton probing
data is given in section~\ref{Section_PRAD}.

In addition
to proton probing, additional plasma measurements 
were obtained using angular-filter refractometry
of an optical probe beam, discussed
in Section~\ref{Section_AFR}.
In addition, surrogate experiments were
fielded at OMEGA which allowed a
Thomson-scattering measureent of plasma 
density ($n_e$) and electron and ion temperatures ($T_e, T_i$),
discussed in section~\ref{Section_TS}.

\subsection{Angular refractometry density measurement}

\label{Section_AFR}

A 263~nm ($4\omega$) optical probe beam was used to probe the plasma density 
evolution from the side-on,
providing imaging data on the evolution of the plasma density.
Optical probing provides data related to plasma density line-integrated
along the beam path; in this experimental setup, the 4$\omega$ beam propagates 
along the $y$-axis, which is along the current sheet.
The $4\omega$ probe beam was 
configured with both angular-filter refractometry (AFR) and shadowgraphy legs.

This section describes the AFR data processing and plasma density results.
For AFR, a mask placed at the Fourier plane produces bright and dark bands in the final image, which 
map the contours of the refractive index of the plasma \cite{HaberbergerPoP2014}.   
This information both displays the qualitative evolution of the plumes and can
be quantitatively processed \cite{AnglandRSI2017}.
AFR \cite{HaberbergerPoP2014}
uses a ``bulls-eye'' mask placed at the optical Fourier plane, which masks certain
refractive angles.
Transitions from light-to-dark and vice-versa correspond to 
fixed values of the plasma refraction angle $|\mathbf{\theta}|$, where
\begin{equation}
\mathbf{\theta} = \frac{1}{2 n_{e,cr,4\omega}} \mathbf{\nabla}_\perp \int n_e d\ell.
\label{EqAFRAngle}
\end{equation}
Qualitatively, the bands correspond to contours of constant $|\nabla n|$.

Both single and colliding plasmas were probed over a series of shots.
We describe two analyses.  First, a 1-D analysis is readily accomplished
along the vertical axis of each plume.  This can be processed to 
directly obtain the line-integrated density as a function of height ($z$),
and this data is a useful support of the reproducibility of the plasmas
from shot to shot.  Second, for single plumes, the entire plume can
be analyzed using a reconstruction procedure and assuming azimuthal symmetry,
which provides a full density map.

\begin{figure}
\centering
\includegraphics{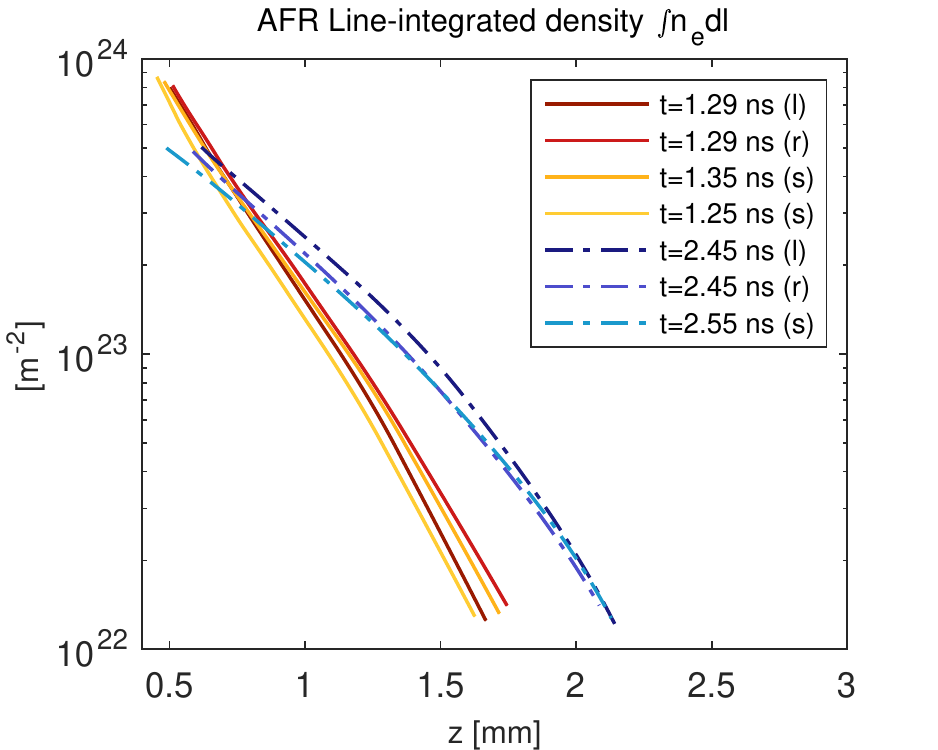}
\caption{AFR Line-integrated density observations for early ($t = 1.25$~ns), and late time ($t = 2.5$~ns), corresponding to
the two time points analyzed extensively in the proton radiography data.  The observation time was repeated
over several shots, both with double-plumes and with single plumes.  
The timing of the 4$\omega$ probe was directly measured for each shot and is indicated in the legend, as
well as whether the shot was a single plume ('s') or was the left ('l') or right ('r') plume
of a double-plume shot.}
\label{FigAFRndl}
\end{figure}

The 1-D analysis provides valuable baseline plasma evolution information
for comparison against simulations and confirming the high reproducibility between
shots.
Along the central (vertical) expansion 
axis of each plume, the density profile is nearly 
1-D ($\partial / \partial z \gg \partial / \partial x = 0$), and
we therefore analyze Eq.~\ref{EqAFRAngle} as a 1-D equation in $z$ along the 
expansion axis.  We locate the AFR transitions in the image, which provide
a locus of points $\{z_j\}$, for the $j$th AFR transition, and where the transition 
refraction angles $\{\theta_j\}$ are known based on the AFR filter construction and calibration \cite{HaberbergerPoP2014}. 
 In between these points, the refraction angle is interpolated in log-space, which is the the logical choice for a quasi-exponential profile, producing
 $\theta(z)$.  Finally,
the line-integrated density is obtained by integrating Eq.~\ref{EqAFRAngle}.
Figure~\ref{FigAFRndl} shows the line-integrated densities along the plume axis
for several shots clustered at times ($t=$~1.3, 2.5 ns) corresponding to the times analyzed in Fig.~\ref{Fig_Prad}.
The plasma profiles are well-reproduced at each time indicating the 
high-reproducibility of the laser performance and plasma production.

\begin{figure}
\centering
\includegraphics[]{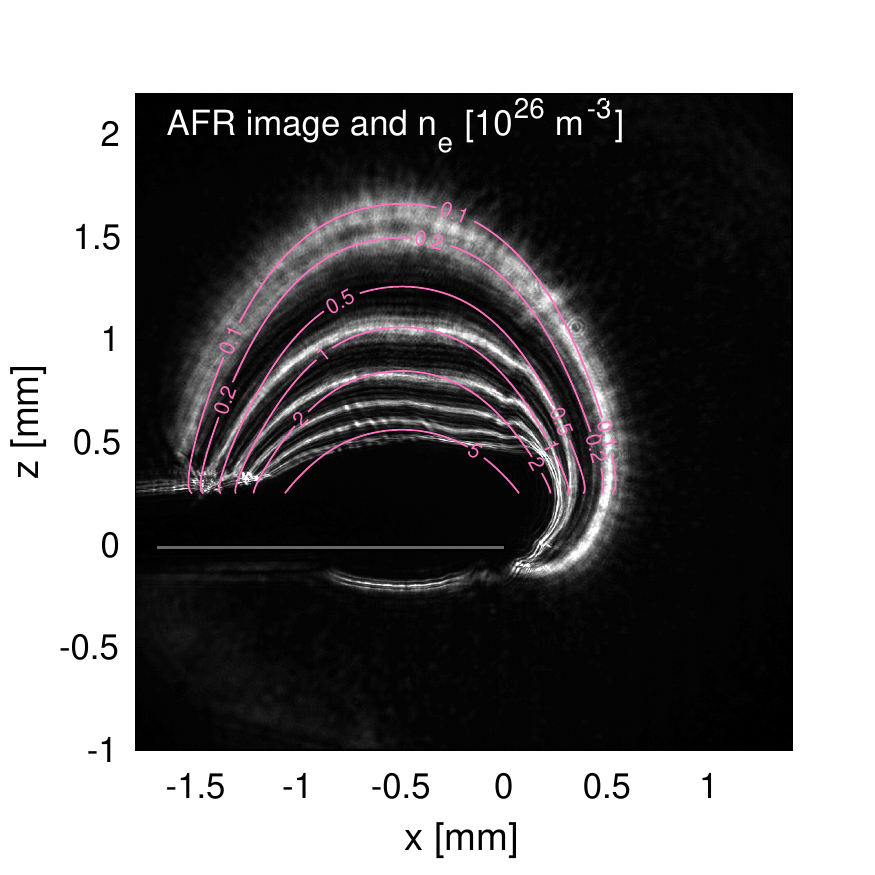}
\caption{Angular filter refractometry image with overplotted
reconstructed density for a single expanding plasma,
$t =$~1.35~ns.  The target foil surface is denoted in gray.
\label{Fig_AFR_wcontour}}
\end{figure}

Secondly, for single-plume shots, a reconstruction method is used to convert the line-integrated 
observations to density, assuming azimuthal symmetry \cite{AnglandRSI2017}.
Single plasmas are the most appropriate for reconstruction 
as they are reasonably azimuthally symmetric.  A typical
AFR image and associated reconstruction are shown in Fig.~\ref{Fig_AFR_wcontour}.
This data is also useful as a comparison against single-point Thomson
scattering measurements below.

\subsection{Thomson scattering}

\label{Section_TS}

Optical Thomson scattering (OTS) was used to provide direct measurements of plasma parameters during the plume interaction.  OTS measurements were conducted on surrogate experiments at OMEGA, using identical targets and drive laser pointing.  Each plume was driven by two overlapped OMEGA 3$\omega$ beams, with on-target laser energy of 950~J per plume, which is only slightly lower than the energies used at OMEGA~EP.  
(Proton radiography was also obtained on these OMEGA shots, and  qualitatively 
confirms the EP proton results presented in this paper,
and which shows that the OMEGA shots were in a common regime.)
A 527~nm ($2\omega$) Thomson scattering probe beam was passed through the plume interaction region.  The scattering volume was directly in the gap between targets, in the plane of the foil, and at a distance $y=~0.8$~mm downstream from the direct midpoint between the plumes, which corresponds to the center of the blue inversion box of Fig.~\ref{Fig_Prad}d. The scattered light was collected and split along two beam lines.  One line passed through a lower resolution (150g/mm) spectrometer to collect spectra associated with electron plasma waves (EPW), while the other line passed through a high resolution (2160g/mm) spectrometer to collect spectra associated with ion acoustic waves (IAW).  Each line was then detected on a separate streak camera with $\sim$50 ps temporal resolution.

The scattered power is proportional to the spectral density function

\begin{equation}
S(\mathbf{k},\omega) = \frac{2\pi}{k}\left|1-\frac{\chi_e}{\epsilon}\right|^2 f_e\left(\frac{\omega}{k}\right) + \frac{2\pi Z}{k}\left|\frac{\chi_e}{\epsilon}\right|^2 f_i\left(\frac{\omega}{k}\right)
\label{eq:ots}
\end{equation}

\noindent where $\omega$ is the frequency, $\mathbf{k=k_s-k_i}$ is the wavevector being probed ($\mathbf{k_i}$ is the wavevector of the incident probe beam and $\mathbf{k_s}$ is the wavevector of the collected light), $Z$ is the ion charge state, $f_e$ and $f_i$ are the electron and ion distribution functions, respectively, $\chi_e$ and $\chi_i$ are the electron and ion susceptibilities, respectively, and $\epsilon=1+\chi_e+\chi_i$ is the longitudinal dielectric function \cite{sheffield_plasma_2011}.  The first term on the RHS of Eq.~\ref{eq:ots} corresponds to the EPW feature, while the second term corresponds to the IAW feature. Assuming Maxwellian distribution functions, plasma parameters can be extracted by iteratively fitting $S(\mathbf{k},\omega)$ to the measured spectra until a best fit is found. Error analysis is done using a Monte Carlo approach, in which the extracted plasma parameters represent the mean value over 50 fits, with the error bars corresponding to the standard deviation \cite{follett_plasma_2016}.

\begin{figure}
\centering
\includegraphics[width=7cm]{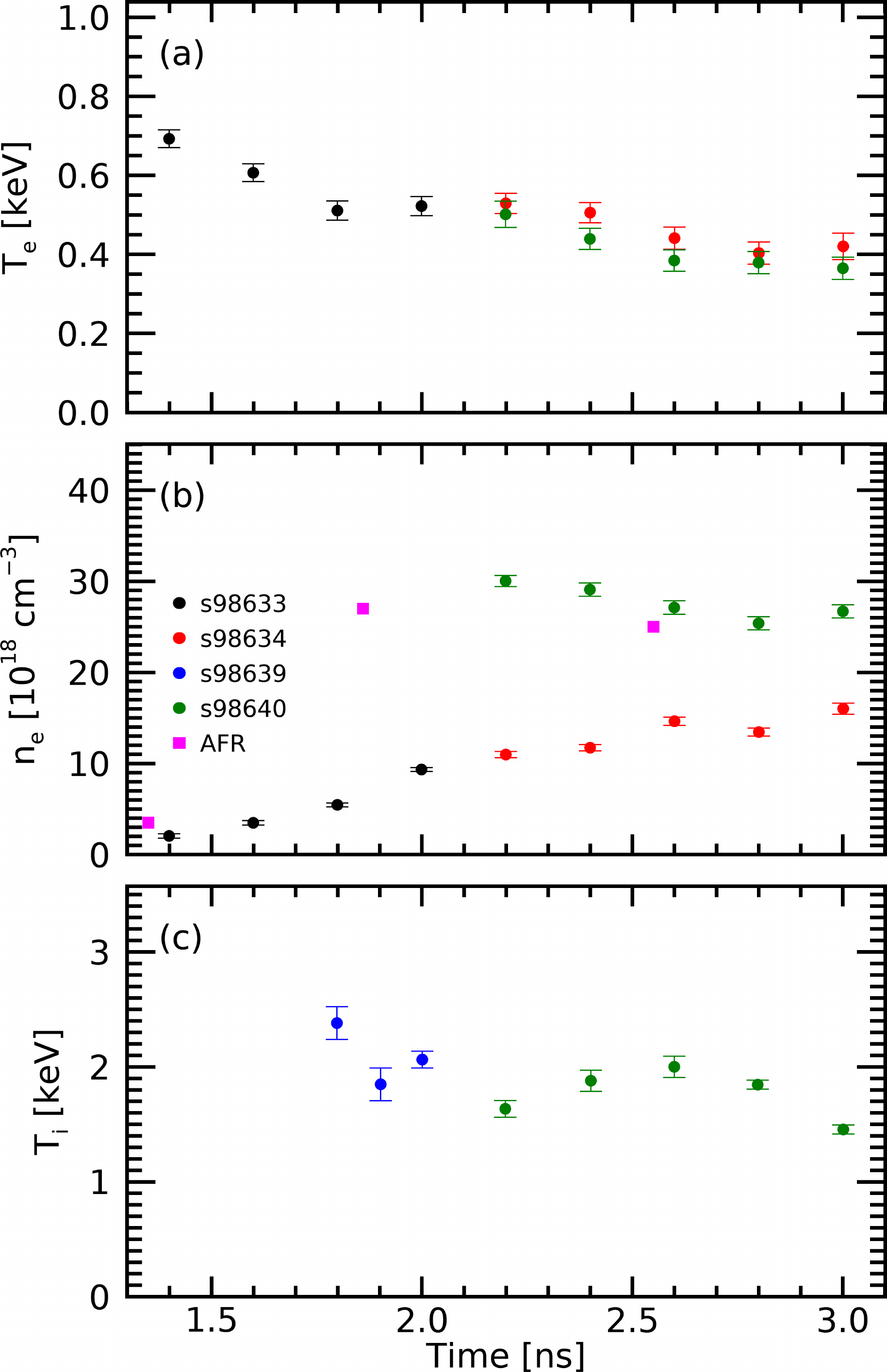}
\caption{Thomson scattering measurements of the reconnection layer.
a) Electron temperature. b) Electron density. c) Ion temperature.  All values derived from analysis of the EPW and IAW spectra.  Also shown in (b) are the electron densities measured with $2 \times $~AFR (magenta squares).}
\label{fig:ots}
\end{figure}

Four shots were analyzed.  All data was collected from the same location, but two probe beam orientations were used (changing the
scattering vector orientation compared to the foils).
The results are shown in Fig.~\ref{fig:ots} for the electron temperature, electron density, and ion temperature.  While the temperature values are consistent between orientations, the density is somewhat higher in one orientation. It is not clear if the difference is due to a relative pointing difference between shots (estimated to be small), or the presence of a density clump in the scattering volume in one of the shots (as the OTS measurement volume is smaller than the magnetic structures apparent in the proton radiography data).  We take the density range as measure of the uncertainty and fold it into the analysis.  We note that many key quantities such as the ion skin depth and \Alfven{} velocity depend relatively weakly on density ($\propto n_e^{-1/2}$), so errors in these quantities are relatively smaller. 

The AFR density measurements were in good agreement with the Thomson scattering data.
The Thomson scattering was obtained at a point in the 
target plane and a distance 0.8~mm in $x$ and 0.8~mm in $y$
compared to the center of the laser foci.
To obtain a similar point from the AFR, we use single
plume AFR data and evaluate at a point 
consistent with radius $r = 0.11$~mm ($= 0.8 \times \sqrt{2}$) and in the foil plane.   We multiple the AFR density by 2 
to account for two plumes, which is in reasonable agreement and supports the TS data (Fig.~\ref{fig:ots}).

\subsection{Plasma parameters}

Table~\ref{Tab_Param} shows estimated plasma parameters in the current sheet
in the region of the blue box of Fig.~\ref{Fig_Prad}d,
and at the  time $t$=2~ns characteristic of the time when instabilities are growing and
 the  current sheet is breaking up.
The upstream magnetic fields and the current sheet width were determined from the
analysis of magenta region in Fig.~\ref{Fig_Prad}b at early time (t = 1.37~ns), where the current sheet is well-formed.
To convert proton radiography observations (line-integrated magnetic fields),
to an average magnetic field, we assumed a relation
$\bar{B} = (1/L_B) \int B d\ell$ to obtain an average magnetic field,
assuming an integration scale length $L_B$.  We assume
a broad range of possible $L_B$ of 0.5--1~mm, 
based by the plasma size as observed in side-on images (Fig.~\ref{Fig_Overview}).  The uncertainty in 
$L_B$ dominates the uncertainty in $\bar{B}$, and we 
fold this into uncertainties in parameters (such as $\beta$ and $S$) and
theory comparisons below.

Plasma density and temperatures are obtained  from Thomson scattering measurements, and we indicate the full
range of values observed between 2.0 and 2.5~ns. 
The full range of values are folded in as uncertainties 
when estimating parameters (such as $S$ and $d_i)$ and
when comparing to theory.
Inflow speeds are estimate based on a 
ballistic ``time-of-flight'' from the plasma birth location,
near the edge of the laser focus at $r = 350~\mu$m relative to each plume center, to the measurement location.

\begin{table}
\centering
\caption{Plasma parameters in reconnection current sheet, blue box of Fig.~\ref{Fig_Prad}d}
\label{Tab_Param}
\begin{tabular}{lr}
\hline\hline
\\
Density $n_e$ (m$^{-3}$) & \sciexp{1--3}{25} \\
Temperature $T_e$ (eV)  & 400--600 \\
Temperature $T_i$ (eV)  & 1600--2400 \\
Inflow speed $V_{in}$ (km/s) & 250 \\
Magnetic field $B$ (T) & 20--40 \\
Current sheet full-length $L_{cs}$ (mm) & 4 \\
Current sheet half-width $a = B_{up} / (dB/dx)$ ($\mu$m) & 30--40 \\
Instability wavelength $\lambda$ ($\mu$m) & 200 \\
Normalized wavenumber $ka$ & 0.9--1.1 \\
$d_i$ (50--50 CH mixture) ($\mu$m) & 60--100\\
$d_e$ ($\mu$m) & 1--1.6 \\
ion-ion MFP (C$^{6+}$--C$^{6+}$, between inflow populations) (mm) & 0.5 \\
\\
Global Lundquist number $S = \mu_0 L V_A / \eta$ & 2500-4500 \\
Current sheet Lundquist number $S_a = \mu_0 a V_A / \eta$ & 20--40 \\

System size $L/d_i$ & 50 \\
$\nu_{ei} / \omega_{ce}$ & 0.04 -- 0.12 \\
$\beta_e = 2 \mu_0 n_e T_e / B^2 $ & 2--8 \\
$\beta_{\mathrm{dyn}} = \mu_0 n_i m_i V_{in}^2 / B^2 $ & 2.5--10\\
electron magnetization $\omega_{ce} / \nu_{ei}$ & 15--50\\
\\
\end{tabular}
\end{table}

\subsection{Proton radiography analysis}

\label{Section_PRAD}

Proton radiographs were obtained on HD-V2 film.
For the qualitative radiography
images presented (Fig.~\ref{Fig_Prad}a-d),
the data was sharpened using an unsharp-mask routine. 
Therefore the images are interpretive and 
do not quantitatively correspond
to a physical proton dose.
However, for all physics analysis (Fig.~\ref{Fig_Prad}e-h),
we start with the clean raw data, and
apply a detailed processing and analysis chain which is now described.

In order to quantitatively analyze the radiographs, 
we use recent results \cite{ChenRSI2016}
on the relation between proton dose and the film exposure,
which shows the relation of the proton fluence (dose) to the film optical depth (OD), obtained
from $OD_{\{RGB\}} = -log_{10} (T_{\{RGB\}})$ 
where $T_{\{RGB\}}$ is the transmission in red, green, or blue
channels reported by the Epson scanner.
The zero-exposure optical depth, $OD_0$, was obtained from
unexposed RCF.  The resulting background was subtracted from the 
$OD$s to produce a signal proportional to proton fluence.  The linearity of $OD$ with
dose has been studied \cite{ChenRSI2016} and is valid for the range of $OD$ we obtain.
We found that applying a small color correction, 
$OD_{R*} = OD_R - \alpha_R OD_B$ and $OD_{G*} = OD_G - \alpha_G OD_B$,
where $\alpha \sim 0.15 - 0.2$ improved the background subtraction
and corrected some long-scale modulations in the film. (The $B$ channel itself has very low response to the proton dose, therefore mainly corrects any OD structuring 
the film substrate).
To test the consistency and robustness of the results, we analyzed 
both $R*$ and $G*$ channels, as well as analyzed 
a several layers near the relevant energy, discussed below.

Proton radiographs are analyzed using a recent reconstruction algorithm \cite{BottJPP2017}.
The algorithm solves for the deflection magnetic field ($\int B_xdz, \int B_y dz$) consistent
with the observed proton fluence pattern.
Previous proton-deflection 
experiments in this plasma parameter regime \cite{PetrassoPRL2009}
showed that the magnetic field is the dominant effect (over
electric fields) in deflecting protons, and therefore
we regard proton deflections as strictly due to magnetic fields.
The algorithm assumes zero-normal-deflection boundary conditions 
($\int \mathbf{B}dz \times \mathbf{n} = 0$, where $\mathbf{n}$ is the 
normal vector through the boundary).
Therefore we chose reconstruction volumes which
extend to $y = \pm 800$~$\mu$m, that is, to the offset of each laser focus.
(This reconstruction volume is larger than the data presented in Fig.~\ref{Fig_Prad}
which has been cropped to focus on the current sheet.)
Based on the structure of the Biermann fields (azimuthal around each
laser focus, though of course this is modified where the plumes interact),
$B_x$ is expected to be small at these $y$ positions.
 
The reconstruction 
algorithm assumes that the ``undisturbed'' proton fluence before reaching the 
plasma is uniform.  However, we found that 
minor changes to the proton fluence, especially at long-wavelength scales, can significantly modify the magnetic field pattern.
This results fundamentally because proton fluence radiographs
are most closely related to $B$-field gradients,
i.e. plasma \textit{current density} \cite{GrazianiRSI2017},
and inferring magnetic fields from this requires a spatial integration
which is sensitive to ``DC'' or long-wavelength offsets.
Furthermore it is clear from examining radiographs in cases with $B=0$ that the
proton beams are not perfectly uniform.
However, the goal in this study is primarily to understand qualitative structures in the plasma \textit{current density},
and our main analysis focuses on this quantity.
To overcome proton fluence non-uniformity 
we devised a high-pass filter which damps
long-scale non-uniformities in the proton signal.
The results in Fig.~\ref{Fig_Prad} use a filter characterized by $FWHM = 600$~$\mu$m.  

To provide an uncertainty and robustness analysis
of this procedure, 
we conducted various robustness and consistency
checks.  That is, we conducted an identical analysis using the 
neighboring layers in the film stack, and
also compared both the $R*$ and $G*$ channels.  The
scan of the neighboring layers only made
a moderate difference in the energy analyzed 
(24.7, 29.2, or 33.2~MeV, i.e. only $\pm~7\%$ in
proton velocity),
but had a larger difference (factor of 2) in the average
proton fluence.  Therefore the comparison of different
layers and color channels is a useful test for uncertainty
related to any film saturation effect, background subtraction,
or any energy-dependent proton-beam aberrations.
Second, we also tested different 
spatial filters applied to the data, from 
FWHM 400~$\mu$m to 800~$\mu$m.
All these tests produced basically indistinguishable results,
confirming that the results are not an artifact of the 
analysis procedure.
As an example metric, the RMS $\delta J / J$, discussed
in relation to Fig.~\ref{Fig_Prad_Zoom}, varied only
$[-10\%, +20\%]$ over all these tests, versus the
factor $\sim$4 change in this quantity from
early to late time.
These confirm the robustness of the results.

\subsection{Tearing Instability Models}

The tearing theory was generalized to anisotropic distribution functions for electrons
and ions in several prior theoretical works \cite{ChenPoF1984, BurkhartPRL1989, QuestPoP2010}.  Anisotropic tearing has attracted 
significant theoretical interest over the years as it is found 
to significantly increase the instability of 
current sheets and may thereby contribute to space physics
phenomena such as magnetospheric substorm onset.

A fundamental parameter is the 
tearing instability index, the dimensionless quantity $a \Delta'$,
which is related to the free magnetic 
energy available to be released
by current sheet breakup.  (Often abbreviated $\Delta'$,  
here we use $a \Delta'$, where $a$ is a length scale related
to the current-sheet profile, to be clear that
we have a dimensionless quantity.).  The tearing index depends
on the current profile and instability wavelength.  For the
canonical current sheet profile with $B_x(y) \propto \tanh(y/a)$, 
and for transverse tearing perturbation $\xi \propto \exp(-i k x + \gamma t)$, it is given by,
\begin{equation}
    a\Delta' = 2(1/ka - ka).
\end{equation}
The quantity can change quantitatively for 
different current profiles, but generally changes sign to negative for $ka \gtrsim 1$,
indicating a stabilization of the mode.  This is a significant
constraint for the present observations where $ka \approx 1$ is observed.

In the fluid regime, resistive
dissipation leads to the classical resistive (Furth-Killeen-Rosenbluth) tearing \cite{FKRPoF1963, BhattacharjeePoP2009}
with growth rate (in the constant-$\Psi$ regime),
\begin{equation}
    \gamma_{R} = 0.96 \frac{V_A}{a} S_a^{-3/5} (a\Delta') (ka)^{2/5}.
    \label{Eq_GammaResTearing}
\end{equation}
Here $S_a$ is the Lundquist number evaluated with the current sheet half-width, 
$S_a = \mu_0 a V_A / \eta$.

The resistive tearing has been generalized to collisionless
regimes \cite{CoppiPRL1966} where electron Landau damping
in the current sheet provides dissipation, and further generalized to
the anisotropic tearing considering pressure anisotropy, where
$T_\perp > T_{||}$ provides additional positive feedback driving instability 
growth \cite{ChenPoF1984, BurkhartPRL1989, QuestPoP2010}.
For anisotropic tearing, we use the analytic formulas of Quest \textit{et al.} \cite{QuestPoP2010} (Eq. (43), 
which were found to be in reasonable agreement with the full-orbit numerical calculations
by the same authors.  Here we use the analytic formulas
because numerical calculations were not conducted in our parameter
regime.  In this theory, the growth rate of anisotropic ion tearing is given by,
\begin{equation}
\gamma_{AT} = k v_{te} \frac{\epsilon_e^{3/2}}{\beta_e} \left\lbrack \frac{\sqrt{\pi}}{2\cdot 2.96} (a\Delta') 
  + \frac{1}{\sqrt{\pi}} \sum_j \frac{\beta_j}{\epsilon_j^{3/2}} A_j \right\rbrack.
  \label{EqAniTearing}
\end{equation}
Here the electron beta $\beta_e = 2 n_e T_e \mu_0 / B^2$, and $\epsilon_e = \rho_e / a$.
One recognizes the 
first term in brackets of Eq.~\ref{EqAniTearing} is the collisionless tearing
drive ($\propto \Delta'$), and the second related to anisotropic ions ($\propto A_j$).
We generalized the treatment to sum over multiple 
ion species ($j$) which may have separate anisotropies $A_j = T_{\perp,\eff,j}/T_{||,j} - 1$,
and with
$\beta_j = 2 \mu_0 n_j T_{\perp,\eff,j}  / B^2$, and $\epsilon_j = \rho_j / a$, where $\rho_j = m_j V_{\perp,j} / Z_j e B$.
The equation was also simplified considering moderate ion anisotropy ($A_j \ll M_i/m_e$), and zero electron anisotropy ($A_e = 0)$.
Quest \textit{et al} 
also introduce a modified expression for 
$a \Delta'$, which we retained, however this term is really only
important for very long wavelengths $ka \ll 1$.
The pure collisionless tearing $\gamma_{CT}$ \cite{CoppiPRL1966}
is obtained from the $A_j = 0$ limit of this equation.
The anisotropic tearing can also be driven by
electron anisotropy, and indeed can provide a stronger lever than
ions and has therefore often
been considered in space physics contexts \cite{KarimabadiGRL2004, QuestPoP2010}.  
However since we infer rapid electron collisional isotropization
in the present experiments, this
effect is difficult to estimate, so we presently just consider the case $A_e = 0$.

Interestingly, the anisotropic tearing obtains a larger growth rate than
pure Weibel instability ($B=0$), for which the growth rate is given by
\cite{DavidsonPoF1972},
\begin{equation}
\gamma_W = \frac{k v_{te}} {\sqrt{\pi}} \left \lbrack \sum_j \frac{(n_j/n_e) Z_j^2}{M_j/m_e} A_j - k^2 d_e^2 \right\rbrack,
\label{Eq_GammaWeibel}
\end{equation}
Here $d_e$ is the electron skin depth.  
For this dispersion relation, we considered the ion-pinch
dispersion relation of Davidson et al, 
in the limit $\gamma / k v_{te} < 1$, and again moderate
ion anisotropy.  
The difference between anisotropic tearing and 
pure Weibel is the presence of the current sheet, which magnetized the electrons, which increases the growth of the instability.  This fact
was pointed out by Quest \textit{et al} \cite{QuestPoP2010}.

We evaluate growth rate of these instabilities for experimental parameters
versus $\lambda = 2\pi / k$.  We consider the typical parameters, 
$B_{up} = 30$~T, $n_e = 2 \cdot 10^{25}$~m$^{-3}$,
and a current sheet half-width $a$ = 35~$\mu$m.  
We consider drive by anisotropic 
C$^6+$ ions, at a fraction 1/7 of the $n_e$ density, with variable
anisotropy $A_i$ from 1--3 indicated in the legend.
For experimental plasma inflows of order 250~km/s (based on time-of-flight)
an anisotropy up to 4 is possible considering the measured $T_i = 2000$~eV.
We plot classical tearing (orange) for unstable $k$ from $S_a^{0.25} < ka < 1$,
which corresponds to unstable constant-$\Psi$ 
modes where the growth-rate of Eq.~\ref{Eq_GammaResTearing} applies.
We also plot collisionless tearing (red), pure Weibel (Eq.~\ref{Eq_GammaWeibel}) (yellow),
and anistropic tearing with $A_i$ = 1--3, (blue through purple).
It is found that classical modes and pure Weibel have too-low growth-rates
and are stable at the observed structure size.
Ion Anisotropy $> 2$ is sufficient to drive the observed instability
growth at $\lambda \approx 200~\mu$m and $\gamma > 10^9$~s$^{-1}$.
The small relative growth of the pure Weibel mode compared to anisotropic tearing
is consistent with the calculations of Quest et al (\cite{QuestPoP2010}, Fig. 8).
Scanning parameters, we find similar curves and the conclusions do not 
change for $B_{up} = $20--40~T and
density $n_e = $~1--3~$\cdot 10^{25}$~m$^{-3}$.

\begin{figure}
    \centering
    \includegraphics{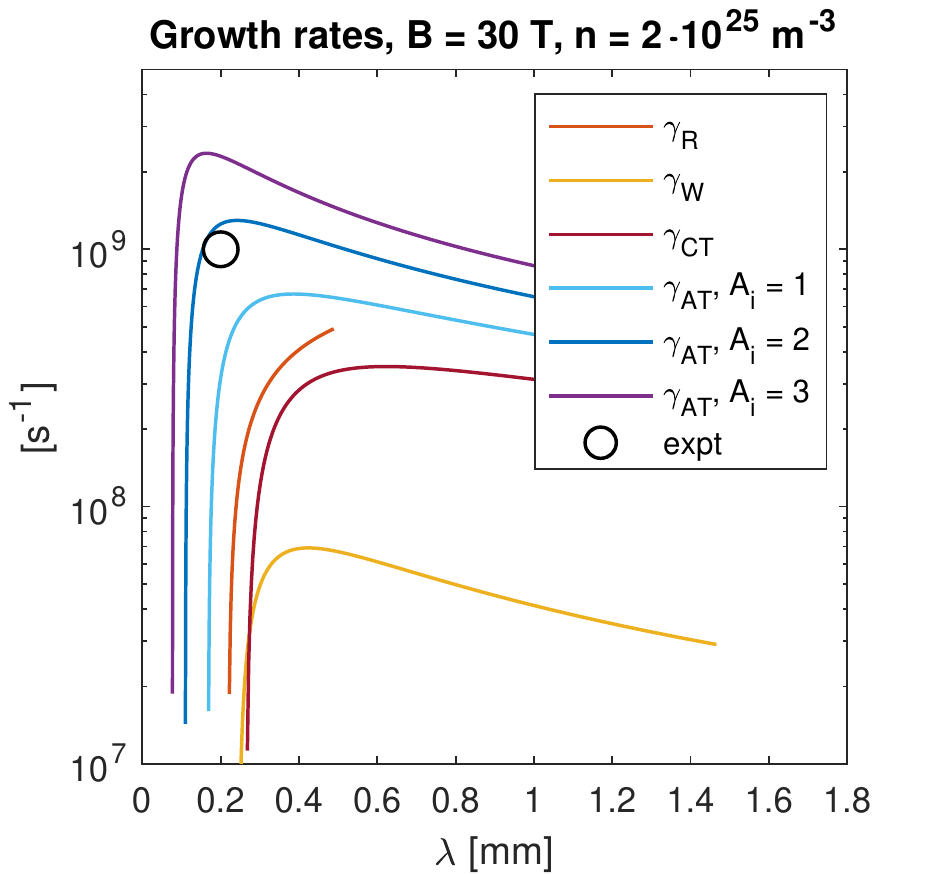}
    \caption{Instability growth calculations for
    various theories under experimental parameters for various wavelength.
    The theory models are described in the text.
    The experimentally-estimated growth rate to generate
    structures at $\lambda = 200~\mu$m is indicated as the circular data point.
    \label{Fig_TearingGrowth}}
\end{figure}

\subsection{Particle-in-cell simulation setup}

The particle-in-cell code PSC \cite{GermaschewskiJCP2016} 
is used to simulate reconnection between colliding laser-plasmas following
the general model which has been used previously \cite{FoxPRL2011, LezhninPoP2018}.
The rich dynamics include compression of opposing magnetic fields 
into an ion-scale current sheet with the simultaneously development
of kinetic instabilities and plasmoid development.
Here we focus on a simulation of 
 current sheet formation and reconnection where
 strong ion anisotropy develops,
to show how this instability can initiate the 
formation of a large number of plasmoids 
with quantitative similarity to the experimental 
observations.    

The simulations are 3-D, but for computational reasons
we simulate a relatively narrow slug from the global geometry.  
Accordingly, the simulations are initialized with two
1-D flows ($n = n(y), V_y = V_y(y), B_x = B_x(y)$, following the 
coordinate system of Fig.~\ref{Fig_Overview}.
The initial profiles are qualitatively similar to used previously \cite{FoxPRL2011, FoxPoP2012b, LezhninPoP2018},
but modified for new DRACO  \cite{RadhaPoP2005} simulations of these gapped experiments.
From these new simulations, the initial profiles are taken at a time after 
magnetic fields have been generated by the Biermann battery effect.
Specifically, the profiles are obtained from a radial cut between the bubbles
at a height $z = 400~\mu$m.
A reference density is obtained based on the peak
plasma density from DRACO, $n_0 = $\sciexp{1.25}{27}~m$^{-3}$,
which defines a global ion skin depth for normalizing the 
simulation box, $d_{i0} \approx 8~\mu$m, for a 50-50~CH mixture.
(For simplicity the PSC simulation itself is conducted with a pure-ion plasma.)
The two plume centers are separated by the global length scale $L_{x0} = 200~d_{i0}$,
making for a very large kinetic simulation, and we use dimensions $25 \times 25~d_{i0}$
in the two transverse directions.
The profile is a smooth gaussian core which transitions to a steeper exponential
to agree with the DRACO profiles, and the 
initial density at the midpoint between the two plumes is $\sim0.015~n_0$.
The plumes are given a uniform initial temperature $T_0$.
Oppositely polarized magnetic fields are initialized in each plume with magnitude $B_0 = 0.05 (n_0 T_0)^{1/2}$.  
Two counter flows are initialized, again per DRACO, with profiles for
each plume $V = V_0 |x| / L_{x0}$, with $V_0 / C_s = 2$, where $C_s$ is the sound speed and $x$ the distance 
from the center of the plume.
The initial condition was modified from previous simulations \cite{FoxPRL2011, FoxPoP2012b, LezhninPoP2018},
to allow ion counterstreaming between the two plumes in the initial
condition. As commonly used with explicit PIC simulations, we use a compressed ion-electron
mass ratio $M_i/m_e = 64$ as well as a compressed speed-of-light, characterized by $m_e c^2 / T_{0} = 25$.

We have found that due to the nature of the  1-D flows driving the simulation, the density
and fields are strongly compressed compared to the
experimental (AFR) measurements of the density.
This is because the flows in the real system compress along the $x$-direction 
while simultaneously de-compressing along the $y$- and $z$-directions 
(vertical and down-current-sheet), whereas the simulation simply compresses in $x$.
This and other 3-D effects were demonstrated in recent work \cite{MatteucciPRL2018} which
simulated the full 3-D evolution and reconnection in these systems with two plumes expanding from a thin foil.
However, those fully-3-D simulations were conducted at a smaller plume separation ($L_{sep} \sim 80 d_{i,ab}$, where $d_{i,ab}$ is the
ion skin depth evaluated at the ablation surface); 
unfortunately the present experimental system is too large ($L_{sep} \sim 400~d_{i,ab}$) to enable
such a simulation with present capabilities.
For this reason, we match simulations to experiments by (1) judicious initialization to 
match dimensionless parameters such as the plasma $\beta_e$ once the current sheet forms; 
and (2), when generating experimental observables such as proton radiography, 
re-scaling lengths to match the local ion-skin depth in the 
current sheet between simulation and experiment, and scaling the magnetic fields 
to the experimental values observed in the current sheet.

\end{document}